\documentclass[11pt]{article}
\usepackage{amsmath}
\usepackage{amsfonts}
\usepackage{amssymb}
\usepackage{amsthm}
\usepackage{tikz}
\usepackage{todonotes}
\usepackage{algorithmic}
\usepackage{url}

\newcommand{\pr}{\operatorname{Prob}}
\newcommand{\maj}{\ensuremath{\operatorname{MAJ}}}

\newcommand{\model}{\ensuremath{\maj_k \circ \maj_k}}
\newcommand{\majn}{\ensuremath{\maj_n}}
\newcommand{\eps}{\varepsilon}

\newcommand{\diff}{\operatorname{diff}}
\newcommand{\walk}{\texttt{walk}}

\newcommand{\circuit}{\ensuremath{C}}
\newcommand{\randcircuit}{\ensuremath{\mathcal{C}}}

\newcommand{\wTCz}{\ensuremath{\widehat{\mathsf{TC}}\vphantom{\mathsf{C}}^0}}
\newcommand{\TC}{\mathsf{TC}}

\usepackage{fullpage}
\usepackage[utf8]{inputenc}
\usepackage[english]{babel}
\usepackage{soul}

\usepackage{calc}
\usepackage{tikz}
\usepackage{enumitem}
\usepackage{amsthm}
\usepackage{amsmath}
\usepackage{array}

\usepackage{amssymb}
\usepackage{comment}
\usepackage{latexsym}
\usepackage{xspace}
\usepackage[hidelinks]{hyperref}

\newtheorem{theorem}{Theorem}
\newtheorem{lemma}[theorem]{Lemma}
\newtheorem{corollary}[theorem]{Corollary}

\begin{document}

\title{Computing Majority
by Constant Depth Majority Circuits
with Low Fan-in Gates
}

\author{
Alexander S. Kulikov\thanks{Steklov Mathematical Institute at St.~Petersburg, Russian Academy of Sciences}
\and Vladimir V. Podolskii\thanks{Steklov Mathematical Institute, Russian Academy of Sciences and
National Research University Higher School of Economics}
}

\date{}

\maketitle
\begin{abstract}
We~study the following computational problem: for which values of~$k$, the majority of~$n$ bits $\maj_n$ can be computed with a depth two formula whose each gate computes a majority function of at most~$k$ bits? The corresponding computational model is denoted by {\model}. We~observe that the minimum value of $k$ for which there exists a {\model} circuit that has high correlation with the majority of $n$ bits  is~equal to $\Theta(n^{1/2})$. We then show that for a randomized {\model} circuit computing the majority of~$n$ input bits with high probability for every input, the minimum value of $k$ is equal to $n^{2/3+o(1)}$. We show a worst case lower bound: if a {\model} circuit computes the majority of $n$ bits correctly on all inputs, then $k\geq n^{13/19+o(1)}$. This lower bound exceeds the optimal value for randomized circuits and thus is unreachable for pure randomized techniques. 
For depth $3$ circuits we show that a~circuit with $k= O(n^{2/3})$ can compute $\maj_n$ correctly on all inputs.
\end{abstract}

\section{Introduction}

In this paper we study majority functions and circuits consisting of them.
These functions and circuits arise for various reasons in many areas of Computational Complexity (see e.g.~\cite{DBLP:books/daglib/0066902,DBLP:books/daglib/0033652,DBLP:books/daglib/0028687}). In particular, the iterated majority function (or recursive majority) consisting of iterated application of majority of small number of variables to itself, turns out to be of great importance, helps in various constructions and provides an example of the function with interesting complexity properties in various models~\cite{DBLP:journals/cc/JuknaRSW99,DBLP:journals/rsa/MagniezNSSTX16,DBLP:journals/rsa/MosselO03,DBLP:journals/siamcomp/KampZ07}.

One of the most prominent examples to illustrate this is the proof by Valiant~\cite{DBLP:journals/jal/Valiant84} that the majority $\maj_n$ of $n$ variables can be computed by a boolean circuit of depth $5.3 \log n$. The construction of Valiant is randomized and there is no deterministic construction known achieving the same (or even reasonably close) depth parameter. The construction works as follows. Consider a uniform boolean formula (that is, tree-like circuit) consisting of $5.3\log n$ interchanging layers of AND and OR gates of fan-in 2. For each input to the circuit substitute a random variable of the function $\maj_n$. Valiant showed that this circuit computes $\maj_n$ with positive probability. Note that AND and OR gates are precisely $\maj_2$ functions with different threshold values. Thus this construction can be viewed as a computation of $\maj_n$ by a circuit consisting of $\maj_2$ gates. There are versions of this construction with the circuits consisting of $\maj_3$ gates (see, e.g.,~\cite{Gold11}).

In this paper we study what happens with this setting if we restrict the depth of the circuit to a~small constant. That is, we study for which $k$ the function $\maj_n$ can be computed by small depth circuit consisting of $\maj_k$ gates. We mostly concentrate on depth $2$ and denote the corresponding model by {\model}.
For example, the majority of $n=7$ bits $x_1, x_2, \dotsc, x_7$ can be computed with the following {\model} circuit for~$k=5$:
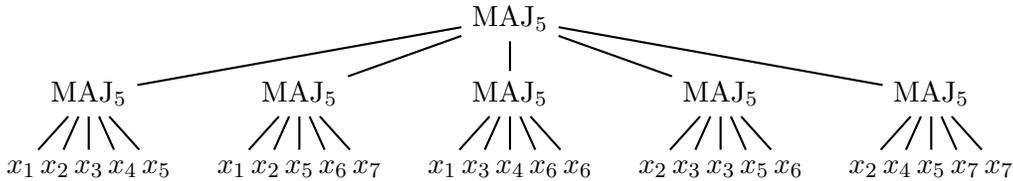
\begin{center}
\begin{tikzpicture}
[thick,
level distance=10mm,
level 1/.style={sibling distance=28mm},
level 2/.style={sibling distance=4.5mm},
]
\node {$\operatorname{MAJ}_5$}
  child{node{$\operatorname{MAJ}_5$} 
    child{node{$x_1$}}child{node{$x_2$}}child{node{$x_3$}}child{node{$x_4$}}child{node{$x_5$}}
  }
  child{node{$\operatorname{MAJ}_5$} 
    child{node{$x_1$}}child{node{$x_2$}}child{node{$x_5$}}child{node{$x_6$}}child{node{$x_7$}}
  }
  child{node{$\operatorname{MAJ}_5$} 
    child{node{$x_1$}}child{node{$x_3$}}child{node{$x_4$}}child{node{$x_6$}}child{node{$x_6$}}
  }
  child{node{$\operatorname{MAJ}_5$} 
    child{node{$x_2$}}child{node{$x_3$}}child{node{$x_3$}}child{node{$x_5$}}child{node{$x_6$}}
  }
  child{node{$\operatorname{MAJ}_5$} 
    child{node{$x_2$}}child{node{$x_4$}}child{node{$x_5$}}child{node{$x_7$}}child{node{$x_7$}}
  }
;
\end{tikzpicture}
\end{center}
We study which upper and lower bounds on $k$ can be shown.

More context to the problem under consideration comes from the studies of boolean circuits of constant depth. The class $\wTCz$ of boolean functions computable by polynomial size constant depth circuits consisting of $\maj$ gates plays one of the central roles in this area. Its natural generalization is the class $\TC^0$ in which instead of $\maj$ gates one can use arbitrary linear threshold gates, that is analogs of the majorities in which variables are summed up with arbitrary integer coefficients and are compared with arbitrary integer threshold. It is known that to express any threshold function it is enough to use exponential size coefficients. To show that $\TC^0$ is actually the same class as $\wTCz$ it is enough to show that any linear threshold function can be computed by constant depth circuit consisting of threshold functions with polynomial-size coefficients (polynomial size can be simulated in $\wTCz$ by repetition of variables). It was shown by Siu and Bruck in~\cite{DBLP:journals/siamdm/SiuB91} that any linear threshold function can be computed by polynomial size depth-3 majority circuit. This result was improved to depth-2 by Goldmann, H{\aa}stad and Razborov in~\cite{DBLP:journals/cc/GoldmannHR92}. More generally, it was shown in~\cite{DBLP:journals/cc/GoldmannHR92} that depth-$d$ polynomial size threshold circuit can be computed by depth-$(d+1)$ polynomial size majority circuit, in particular establishing the class of depth-$2$ threshold circuits as one of the weakest classes for which we currently do not know superpolynomial size lower bounds. The best lower bound known so far is $\Omega({\frac{n^{3/2}}{\log^3 n}})$ by Kane and Williams~\cite{DBLP:conf/stoc/KaneW16}.

Note, however, that the result of~\cite{DBLP:journals/cc/GoldmannHR92} does not translate to monotone setting. Hofmeister in~\cite{DBLP:conf/coco/Hofmeister92} showed that there is a monotone linear threshold function requiring exponential size depth-2 monotone majority circuit. Recently this result was extended by Chen, Oliveira and Servedio~\cite{DBLP:journals/eccc/ChenOS15} to monotone majority circuits of arbitrary constant depth.

Our setting can be viewed as a scale down of the setting of~\cite{DBLP:journals/cc/GoldmannHR92} and~\cite{DBLP:conf/coco/Hofmeister92}. In~\cite{DBLP:journals/cc/GoldmannHR92,DBLP:conf/coco/Hofmeister92} exponential weight threshold functions are compared to depth-2 threshold circuits with polynomial weights. In our setting we compare weight-$n$ threshold functions with depth-2 threshold circuits with weights $k$. In this paper we consider monotone setting.

Another context to our studies comes from the studies of lower bounds against $\wTCz$. Allender and Kouck{\'y} in~\cite{DBLP:journals/jacm/AllenderK10} showed that to prove that some function is not in $\wTCz$ it is enough to show that some self-reducible function requires circuit-size at least $n^{1+\eps}$ when computed by constant depth majority circuit. As an intermediate result they show that $\maj_n$ can be computed by $O(1)$-depth circuit consisting of $\maj_{n^\eps}$ gates and of size $O(n\log n)$.
This setting is similar to ours, however in this paper we are interested in the precise depth and we do not pose additional bounds on the size of the circuit (however note that the bound on the fan-in $k$ of the gates and the bound on the depth $d$ of the circuit naturally imply the bound of $O(k^d)$ on the size of the circuit). 

We consider three models of computation of the majority function: computation on most of the inputs (that is, high correlation with the function), randomized computation with small error probability on all inputs, and deterministic computation with no errors. We prove the following lower and upper bounds for our setting.
\begin{itemize}
\item {\em Circuits with high correlation.} We observe that  the minimum value of $k$ for which there exists a {\model} circuit that computes $\maj_n$ correctly on 2/3 fraction of all the inputs, is equal to $\Theta(n^{1/2})$. A~lower bound is proved by observing that a circuit with $k=\alpha n^2$ does not even have a possibility to read a large fraction of input bits when the constant $\alpha$ is small enough. We show that in this case the circuit errs on many inputs. An~upper bound is proved for the following natural circuit: pick $k=\Theta(n^{1/2})$ random subsets of the $n$ inputs bits of size~$k$, compute the majority for each of them, and then compute the majority of results. Such a circuit computes $\maj_n$ correctly with high probability on inputs whose weight is not too close to~$n/2$. By~tuning the parameters appropriately, we ensure that the middle layers of the boolean hypercube (containing inputs where the circuits errs with high probability) constitute only a small fraction of all the inputs.
\item {\em Randomized circuits.} We prove that for a probabilistic distribution {\randcircuit} of {\model} circuits with a property that for every input $A \in \{0,1\}^n$ the probability that ${\randcircuit}(A)={\maj_n}(A)$ is $1-\eps$ for a constant $\eps>0$, the minimum value of $k$ is $n^{2/3}$, up~to polylogarithmic factors. A~lower bound is proved by showing that a small circuit must err on a large fraction of minterms/maxterms of $\maj_n$. Roughly, the majority function have many inputs $A \in \{0,1\}^n$ with a property that changing a single bit in~$A$ changes the value of the function (these are precisely minterms and maxterms of $\maj_n$). If $k$ is small enough, a~{\model} circuit can reflect such a change in the value only for a small fraction of inputs. To show an upper bound, we split the $n$ input bits into blocks and for each block compute several middle layers values of the bits of this block in sorted order. We then compute the majority of all the resulting values. We show that by tuning the parameters appropriately, one can ensure that this circuit err only on a polynomially small fraction of inputs.   
\item {\em Deterministic circuits.} The trivial upper bound on $k$ is $k\leq n$. We do not have any nontrivial upper bound on $k$ for depth 2 circuits. We however have examples for $n=7,\ 9,\ 11$ of circuits with $k=n-2$. For depth $3$ we have an upper bound $O(n^{2/3})$ which coincides with the optimal value for depth $2$ randomized circuits up to polylogarithmic factor. We prove this upper bound by extending the construction of upper bound for depth 2 randomized circuits. We use an extra layer of the circuit to preorder the inputs. Regarding the lower bound for depth 2 we observe that the following simple special case cannot compute $\maj_n$: each gate is a standard majority (that is, with threshold $k/2$) of exactly $k=n-2$ distinct variables. Next, we proceed to the main result of the paper. We show that the minimum value of $k$ for which there is a depth 2 circuit computing $\maj_n$ on all inputs is at least $n^{13/19}$ up to a polylogarithmic factor. 

Note that this lower bound exceeds the optimal value of $k$ for randomized circuits. Thus, despite the fact that randomized techniques is extensively used for studying majority and circuits constructed from it and proves to be very powerful (recall for example Valiant's result~\cite{DBLP:journals/jal/Valiant84}), in our setting using combinatorial methods we prove a~lower bound that is unreachable for a pure probabilistic approach. The proof of this result however is still probabilistic: in essence we consider a circuit with $k$ smaller than $n^{13/19}$ and build a distribution on inputs that fools this circuit. The catch is that the distribution is tailored to fool this particular circuit: it is constructed via a non-trivial process that involves the values of the gates of the circuit on various inputs.
\end{itemize}

The rest of the paper is organized as follows.
In Section~\ref{sec:prelim} we give necessary definitions and collect technical statements.
In Section~\ref{sec:high-correlation} we study circuits computing the function with high correlation.
In Section~\ref{sec:randomized} we give bounds for randomized circuits.
In Section~\ref{sec:deterministic} we study deterministic circuits.
Finally, in Section~\ref{sec:conclusion} we give concluding remarks and state several open problems. 
Most of the proofs are moved from the main text to Appendix.

\section{Definitions and Preliminaries} \label{sec:prelim}
In this section we will give necessary definitions and collect technical statements that we will use throughout the paper. 

We are going to study circuits computing the well known boolean majority function defined as follows:
\(\maj_n(x_1, x_2, \dotsc, x_n) = [\sum_{i=1}^{n}x_i \ge n/2] \). Here, $[\cdot]$ denotes the standard Iverson bracket: for a predicate $P$, $[P]=1$ if $P$ is true, and $[P]=0$ is $P$ is false. To abuse notation, we will also use $[m]$ to denote the set $\{1,2,\dotsc,m\}$.

It will be convenient to use $X=\{x_1,x_2,\dotsc,x_n\}$ for the set of $n$ input bits. For an assignment $A \colon X \to \{0,1\}$, by $w(A)$ we denote the weight of~$A$, that is, $\sum_{x \in X}A(x)$. For a subset of input variables $S \subseteq X$, by $w_S(A)$ we denote the weight of~$A$ on~$X$: $w_S(A) = \sum_{x \in S}A(x)$. By $\maj_S(X)$ we denote the majority function on~$S$: $\maj_S(X)=[\sum_{x \in S}x \ge |S|/2]$. In particular, $\maj_X$ is just $\maj_n$.

An assignment $A \colon X \to \{0,1\}$ is called a minterm of $\maj_n$ if $\maj_n(A)=1$, but flipping any $1$ to $0$ in $A$ results in an assignment $A'$ such that $\maj_n(A')=0$. A~maxterm is defined similarly with the roles of~$0$ and~$1$ interchanged.

The majority function is a special case of a threshold function: $f(X)=[\sum_{i=1}^{n} a_ix_i \ge t]$.
For such a function $f$ and an assignment $A \colon X \to \{0,1\}$, let difference of $f$ w.r.t. $A$ be \(\diff(f, A)=\sum_{i=1}^{n} a_iA(x_i)-t\). In particular, $f(A)=1$ iff $\diff(f,A) \ge 0$. 

The {\model} computational model that we study in this paper is defined as a depth two formula (we will call it a circuit also) consisting of arbitrary {\em threshold} gates of the form $[\sum{c_ix_i} \ge t]$ where $c_i$'s are positive integers (this, in particular, means that the model is monotone) and $\sum c_i \le k$. At the same time, abusing notation, by $\maj_n$ and $\maj_X$ we always mean the standard majority function. We note that the coefficients in $c_i$ can be simulated by repetition of variables (note that $k$ upper bounds the sum of the coefficients). So the generalization of the $\maj_k$ in the circuit compared to $\maj_n$ is that we allow arbitrary threshold. We note however, that if we are interested in the value of $k$ up to a constant factor (which we usually do), it is not an actual generalization since any threshold can be simulated by substituting constants $0$ and $1$ as inputs to the circuit.

For a gate $G$ at the bottom level of a {\model} circuit, by $X(G)$ we denote the set of its input bits.

\subsection{Tail Bounds and Binomial Coefficients Estimates}

We will use the following versions of Chernoff--Hoeffding bound (see, e.g.,~\cite{DBLP:books/daglib/0025902}).
\begin{lemma}[Chernoff--Hoeffding bound] \label{lem:chernoff}
Let $Y =\sum_{i=1}^m Y_i$, where $Y_i$, $i \in [m]$, are independently distributed in $[0, 1]$. Then for all $t > 0$,
\[
\Pr[Y > E[Y] + t], \Pr[Y < E[Y] - t] \leq e^{-2t^2/m}.
\]
For all $\eps>0$
\[
\Pr[Y > (1+\eps)E[Y]], \Pr[Y < (1-\eps)E[Y]] \leq e^{-\frac{\eps^2}{3}E[Y]}.
\]
\end{lemma}

We will also need the following well known estimates for the binomial coefficients (see, e.g.,~\cite[Section~4.2]{DBLP:books/daglib/0090945}):
\begin{lemma} \label{lem:binomials}
The middle binomial coefficient is about $n^{1/2}$ times smaller than $2^n$. To make it smaller than $2^n$ by arbitrary polynomial factor, it is enough to step away from the middle by about $\Theta(\sqrt{n\ln n})$ ($0<c<1$ is a constant below):
\begin{equation}\label{eq:movemid}
{n \choose n/2}=\Theta(1)\cdot 2^n \cdot n^{-1/2} \text{\quad and \quad}
{n \choose \frac n2+\frac{c\sqrt{n\ln n}}{2}} = \Theta(2^n n^{-\frac 12}n^{-\frac{c^2}{2}}) \, .
\end{equation}
\end{lemma}

\subsection{Hypergeometric Distribution}

The hypergeometric distribution is defined in the following way.
Consider a set $S$ of size $m$ and its subset $S'$ of size $k$. Select (uniformly) a random subset $T$ of size $t$ in~$S$. Then a random variable $|T \cap S'|$ has a hypergeometric distribution. The values $m$, $k$ and $t$ are parameters here.
We will need the following basic properties of this distribution.
For the sake of completeness their proofs can be found in the Appendix (Section~\ref{sec:tech}).

\begin{lemma} \label{lem:wide-layers}
Suppose in hypergeometric distribution $k=k(m)\leq m/2$ (that is, $k$ may depend on $m$). Let $t=t(m)$ be a function with $\eps m< t < (1-\eps)m$ for some constant $0<\eps<1$. Then, for any integer $l$, $\pr(|T \cap S'|=l)=O(k^{-1/2})$, where $O(\cdot)$ is for $m\rightarrow \infty$ and the constant inside $O(\cdot)$ depends on $\eps$, but does not depend on $m$, $k$ and $t$.
Moreover, if $|l-\frac{tk}{m}|=O(1)$, then this probability is in fact $\Theta(k^{-1/2})$.
\end{lemma}

\begin{lemma} \label{lem:wide-layers-general}
Suppose in hypergeometric distribution $k=k(m)\leq m/2$ (that is, $k$ may depend on $m$). Let $t=t(m)$ be a function with $\eps m< t < (1-\eps)m$ for some constant $0<\eps<1$. Consider an arbitrary antichain $A$ on~$S'$ (that is, a~family of subsets of $S'$ none of which is a subset of some other). Then the probability $\Pr[T\cap S\subseteq A] = O(k^{-1/2})$, where $O(\cdot)$ is for $m\rightarrow \infty$ and the constant inside $O(\cdot)$ depends on $\eps$, but does not depend on $m$, $k$ and $t$.
\end{lemma}

\begin{lemma}\label{lem:hypergeo-simple}
For $S$, $S'$ and $T$ as above we have
\(\pr\{|T \cap S'| \ge l\} \le (tk/m)^l \, .\)
\end{lemma}

\section{Circuits with High Correlation} \label{sec:high-correlation}
In this section, we prove that the minimum value of $k$ for which there exists a {\model} circuit that computes $\maj_n$ correctly on, say, 2/3 fraction of all the inputs, is equal to $\Theta(n^{1/2})$.

\subsection{Upper Bound}

\begin{theorem}
\label{thm:correlation-upper}
For any $\varepsilon>0$, there exists a circuit $\circuit$ in {\model}, where $k=O_{\varepsilon}(n^{1/2})$, that agrees with $\maj_n$ on at least $(1-\varepsilon)$ fraction of the boolean hypercube $\{0,1\}^n$.
\end{theorem}
\begin{proof}[Proof Sketch]
The required circuit is straightforward: we just pick $k$ random subsets $S_1, S_2, \dotsc, S_k$ of $X$ of size $k$, compute the majority for each of them, and then compute the majority of the results:
\({\circuit}(X)=\maj_k(\maj_{S_1}(X), \maj_{S_2}(X), \dotsc, \maj_{S_k}(X)) \, .\)
The resulting circuit has a high probability of error on middle layers of the boolean hypercube. We however select the parameters so that all the inputs from these middle layers constitute only a small $\varepsilon/2$ fraction. We  then show that 
among all the remaining inputs (not belonging to middle layers) there is only a fraction $\varepsilon/2$ (of all the inputs) where $\maj_n$ may be computed incorrectly. Overall, this gives a circuit that errs on at most $\varepsilon$ fraction of the inputs. A~detailed proof is provided in Section~\ref{app:corr} in the Appendix.
\end{proof}

\subsection{Lower Bound}

Next we show that this upper bound is tight.

\begin{theorem}\label{lem:corrlow}
Let ${\circuit}$ be a {\model} circuit that computes $\maj_n$ correctly on a fraction $1-\varepsilon$ of all $2^n$ inputs for a constant $\epsilon \le 1/3$. Then $k=\Omega_\varepsilon(n^{1/2})$.
\end{theorem}

\begin{proof}[Proof Sketch]
Let $k=\alpha n^{1/2}$ for a small enough constant $\alpha=\alpha(\varepsilon)$. Note that such a circuit can read at most $k^2=\alpha^2n$ of the input bits. This means that the circuit errs on a large number of inputs. All formal estimates are given in Section~\ref{app:corr} in the Appendix.
\end{proof}

\section{Randomized Circuits} \label{sec:randomized}
The upper bound from the previous section, however, is not enough to obtain a randomized circuit since the construction in Theorem~\ref{thm:correlation-upper} has a very high error probability on the middle layers of the boolean cube. By a randomized circuit here we mean a probabilistic distribution on deterministic circuits computing the function correctly on every input with high probability.

It is not difficult to see that the existence of a randomized circuit is equivalent to an existence of a deterministic circuit computing the function correctly on most of minterms and maxterms (the proof of the following lemma can be found in Section~\ref{app:randc-proof} in the Appendix).

\begin{lemma}\label{lem:randc}
If there exists a randomized circuit {\randcircuit} in {\model} computing $\maj_n$ with error probability~$\eps$, then there exists a deterministic circuit {\circuit} in {\model} computing $\maj_n$ incorrectly on at most~$\eps$ fraction of minterms and  maxterms. Conversely, if there exists a deterministic circuit {\circuit} in {\model} computing $\maj_n$ incorrectly on at most $\eps$ fraction of minterms and  maxterms, then there exists a randomized circuit {\randcircuit} in {\model} computing $\maj_n$ with error probability at most $2\eps$.
\end{lemma}

So from now on instead of probabilistic circuits we study deterministic circuits with high accuracy on two middle layers of $\{0,1\}^n$.

\subsection{Upper Bound}
\begin{theorem}\label{thm:randupper}
There exists a randomized {\model} circuit computing $\maj_n$ incorrectly on each input with probability at most $1/\operatorname{poly}(n)$ for $k=O(n^{2/3}\log^{1/2}n)$.
\end{theorem}

\begin{proof}[Proof Sketch]
Partition the set of $n$ input bits into $n^{1/3}$ blocks of size $p=n^{2/3}$: $X=X_1 \sqcup X_2 \sqcup \dotsc \sqcup X_{\frac np}$. For each block $X_i$, compute $[\sum_{x \in X_i}x \ge m]$ for all $m \in [\frac p2 - \frac t2, \frac p2 + \frac t2]$ for $t \approx n^{1/3}\log^{1/2}n$, and return the majority of results. By~selecting the right value of $t$, this gives a circuit that computes $\maj_n$ incorrectly only on a fraction $\frac{1}{\operatorname{poly}(n)}$ of inputs. The detailed proof is given in Section~\ref{app:randc-proof} in Appendix.
\end{proof}

\subsection{Lower Bound}

In this subsection we show that the upper bound of the previous subsection is essentially tight.

\begin{theorem}\label{thm:randlow}
If a {\model} circuit computes $\maj_n$ on a $1-\eps$ fraction of minterms and maxterms for $\eps < 1/10$, then $k=\Omega(n^{2/3})$.
\end{theorem}

\begin{proof}[Proof Sketch]
The majority function have many inputs $A \in \{0,1\}^n$ with a property that changing a single bit in~$A$ changes the value of the function (these are precisely minterms and maxterms of $\maj_n$). If $k=\alpha n^{2/3}$ for a small enough constant $\alpha$, a~{\model} circuit can reflect such a change in the value only for a small fraction of inputs. A~detailed proof is given in Section~\ref{app:randc-proof} in the Appendix.
\end{proof}

\section{Deterministic Circuits} \label{sec:deterministic}

In this section, we consider {\model} circuits that compute $\maj_n$ correctly on all $2^n$ inputs.

\subsection{Upper Bounds}

\subsubsection{Depth Two}
In this section, we present {\model} circuits computing $\maj_n$ on all inputs for $k=n-2$ when $n=7,9,11$. These circuits were found by extensive computer experiments (with the help of SAT-solvers). Though the examples below look quite ``structured'', currently, we do not know how to generalize them to all values of~$n$ (not to say about constructing such circuits for sublinear values of~$k$). In the examples below, we provide $k=n-2$ sequences consisting of $k=n-2$ integers from $[n]$. These are exactly the input bits of the $k$ majority gates at the lower level of the circuit. That is, each gate computes the standard $\maj_k$ function (whose threshold value is $k/2$).\\[3mm]
\begin{minipage}[t]{0.15\textwidth}
$n=7$:
\begin{verbatim}
1 2 3 4 5
1 2 3 6 7
1 4 5 6 7
2 2 4 5 6
3 4 5 7 7
\end{verbatim}
\end{minipage}
\null\hfill
\begin{minipage}[t]{0.2\textwidth}
$n=9$:
\begin{verbatim}
1 2 3 4 5 6 7
1 2 3 4 5 8 9
1 2 3 6 7 8 9
1 4 5 6 7 8 9
1 3 5 5 7 9 9
1 2 4 6 6 8 8
2 3 4 5 6 7 8
\end{verbatim}
\end{minipage}
\null\hfill
\begin{minipage}[t]{0.25\textwidth}
$n=11$:
\begin{verbatim}
1 2 3 4 5 6 7  8  9
1 2 3 4 5 6 7 10 11
1 2 3 4 5 8 9 10 11
1 2 3 6 7 8 9 10 11
1 4 5 6 7 8 9 10 11
1 2 2 4 6 6 8 10 10
2 4 4 5 6 7 8 10 11
3 3 5 5 7 7 8  9 11
3 3 6 8 9 9 9 10 10
\end{verbatim}
\end{minipage}\\[3mm]

Note that in the examples above there is always a gate in the circuit having one variable repeated more than once. Next we observe that this is unavoidable for $k=n-2$.

\begin{lemma} \label{lem:n-2-lower} For odd $n$ there is no $\model$ circuit for $k=n-2$ with all gates being standard majorities (that is, with the threshold $n/2$) and having exactly $k$ distinct variables in each gate on the bottom level.
\end{lemma}
We provide a proof of this lemma in Section~\ref{app:det} in the Appendix.
\subsubsection{Depth Three}

In this section we extend the proof of the upper bound for randomized depth-2 circuits (Theorem~\ref{thm:randupper}) to construct a circuit of depth 3 for $k = O(n^{2/3})$ computing majority on all inputs.

\begin{theorem}\label{thm:depth-3-upper}
For $k=O(n^{2/3})$ there is a circuit of depth $3$ computing majority of $n$ variables on all inputs.
\end{theorem}

\begin{proof}[Proof Sketch]
We adopt the strategy of the proof of Theorem~\ref{thm:randupper}. That is, we break inputs into $O(n^{1/3})$ blocks, compute majorities on each block on middle $O(n^{1/3})$ layers and then compute the majority of the results. We use the third layer of majority gates to induce additional structure on the inputs. The full proof is given in Section~\ref{app:det} in the Appendix.
\end{proof}

\subsection{Lower Bound}

In this section we will extend the lower bound on $k$ above $\Omega(n^{2/3})$ for depth-2 circuits computing $\maj_n$ on all inputs.

\begin{theorem} \label{thm:worst-case-lower-general}
Suppose a {\model} circuit computes $\maj_n$ on all inputs. Then 
\(
k = \Omega\left(n^{13/19}\cdot (\log n)^{-2/19}\right).
\)
\end{theorem}

We also show the following result for the special case of circuits with bounded weights.
\begin{theorem} \label{thm:worst-case-lower-weighted}
Suppose a {\model} circuit computes $\maj_n$ on all inputs and uses only weights at most $W$ in the gates. Then 
\(
k = \Omega(n^{7/10}\cdot(\log n)^{-1/5}\cdot W^{-3/10}) \, .
\)
\end{theorem}

In particular, we get the following corollary for circuits with unweighted gates.

\begin{corollary} \label{cor:worst-case-lower-unweighted}
Suppose a {\model} circuit computes $\maj_n$ on all inputs and each variable occurs in each gate of the bottom level at most once. Then 
\(
k = \Omega(n^{7/10}\cdot (\log n)^{-1/5}) \, .
\)
\end{corollary}

The rest of this section is devoted to the unified proof of these lower bounds. To follow this proof it is convenient to think that $k=n^{\frac 23+\varepsilon}$ for some small $\eps>0$. In the end it will indeed be the case  up to a logarithmic factor. In the proof we will calculate everything precisely in terms of parameters~$n$ and~$k$, but we will provide estimates assuming that $k=n^{2/3+\eps}$. This is done in order to help the reader to follow the proof.

Let $F$ be a {\model} formula computing $\majn$ on all inputs from $\{0,1\}^n$. Denote by $W$ the largest weight of a variable in gates of~$F$.

\subsubsection{Normalizing a formula}

We start by ``normalizing''~$F$, that is, removing some pathological gates from~$F$. We do this in two consecutive stages. 

{\em Stage 1: removing AND-like gates.} We will need that no gate can be fixed to 0 by assigning a~small number of variables to 0 (here and in what follows we consider gates from the bottom level only). For this, assume that there is a gate that can be fixed to 0 by assigning to 0 less than $n/(100k)=n^{1/3-\eps}/100$ variables. Take these variables and substitute them by 0; this kills this gate (and might potentially introduce new gates with the property). 
We repeat this process until there are no bad gates left. Recall that the number of gates at the bottom level is at most $k = n^{2/3 + \eps}$, so there are at most $k = n^{2/3+\eps}$ steps in this process and hence $n$ is replaced by $99n/100$.  To simplify the presentation, we just assume that $|X|=n$ and that $F$ has no bad gates.

{\em Stage 2: removing other pathological gates and variables.} The formula $F$ contains at most $k^2=n^{\frac 43+2\eps}$ occurrences of variables (counting with multiplicities). Let $x^* \in X$ be a least frequent variable at the leaves. The number of occurrences of $x^*$ is at most $k^2/n = n^{1/3+2\eps}$. In the following we consider only assignments $A$ with $\diff(\majn, A)=-1$ setting $x^*$ to~$0$: 
\[ \mathcal{A}^*=\{A \colon X \to \{0,1\} \mid \diff(\majn, A)=-1\text{ and }A(x^*)=0\}\,. \] 
We also focus on the gates from the first level that depend on $x^*$, denote this set by $\mathcal{G}^*$ (hence $|\mathcal{G}^*| \le k^2/n = n^{1/3+2\eps}$). 
The total number of variables in the gates from $\mathcal{G}^*$ (counting with multiplicities) is at most $k |\mathcal{G}^*| \le k^3/n =  n^{1+3\eps}$.

We now additionally normalize the circuit. We get rid of the following bad gates and variables:
\begin{enumerate}
\item gates in $\mathcal{G}^*$ that can be assigned to 1 by fixing less than $n^2/(100k^2) = n^{2/3-2\eps}/100$ variables in $X\setminus\{x^*\}$ to 1;
\item gates in $\mathcal{G}^*$ with the weight of the variable $x^*$ greater than $100k^3/n^2 = 100n^{3\eps}$;
\item variables with total weight in all gates in $\mathcal{G}^*$ greater than $100k^3/n^2 = 100 n^{3\eps}$.
\end{enumerate}
We do this by the following iterative procedure. If on some step we have a gate violating 1 we fix less than $n^2/(100k^2) = n^{2/3-2\eps}/100$ variables of the gate among $X\setminus\{x^*\}$ to 1 to assign the gate to a constant. If we have a gate violating 2 we fix all the variables of the gate among $X\setminus\{x^*\}$ to 1 to assign the gate to a constant. If we have a variable violating 3, we fix the violating variable to~1.

We note that if we fix all variables in $G \in \mathcal{G}^*$ except $x^*$ to 1, then the gate becomes constant. Indeed, if it is not constant, then the gate outputs $0$ on the input with $x^*=0$ and the rest of the variables equal to 1. Due to the monotonicity of the gate this means that the gate can be assigned to $0$ by assigning a single variable $x^*$ to 0 and we got rid of the gates with this property on the first stage of the normalization.

Since there are at most $k^2/n = n^{1/3+2\eps}$ gates in $\mathcal{G}^*$ we will fix at most $n/100$ variables for case 1. Since the total weight of $x^*$ is at most $k^2/n = n^{1/3+2\eps}$ we will have case 2 at most $n/(100k) = n^{1/3-\eps}/100$ times. Since each gate has at most $k = n^{2/3+\eps}$ variables we will fix at most $n/100$ variables for the second case. Since the total weight of all variables in $\mathcal{G}^*$ is at most $k^3/n = n^{1+3\eps}$ we will fix at most $n/100$ of them for the case 3.

In particular, we have fixed all variables having weight greater than $100k^3/n^2 = 100n^{3\eps}$ in some gate of $\mathcal{G}^*$, so from now on we can assume that $W \leq 100k^3/n^2$.

Another important observation is that now in each gate there are at least $n^2/(100k^2)$ inputs. Otherwise the gate falls under condition of case 1 above.

After this normalization $n$ is replaced by $97n/100$.  To simplify the presentation, again, we assume that $|X|=n$ and the circuit $F$ is normalized.
Note that after redefining $n$ the threshold of the function $\majn$ we are computing is no longer $n/2$, but rather is $cn$ for some constant $c$ close to $1/2$. This does not affect the computations in the further proof.

\subsubsection{Analysis}
The key idea is that if we have an assignment $A \in \mathcal{A}^*$ with $\diff(\maj_n,A)=-1$, then there is a gate $G \in \mathcal{G}^*$ with $-W \le \diff(G,A) \leq -1$. Indeed, otherwise we can flip the variable $x^*$, the value of $\maj_n$ changes, but none of the gates changes their value. 
The plan of the proof is to construct an assignment that violates this condition. This will lead to a contradiction.

For an assignment $A \in \mathcal{A}^*$ with $\diff(\maj_n,A)=-1$ and integer parameters $s$ and $d$ (to be chosen later), consider the following process $\walk(A, s, d)$.

\begin{algorithmic}[1]
\STATE $A_0 \gets A$\\
\FOR{$i=1$ to $s$}
\IF{for each $G \in \mathcal{G}^*$, $\diff(G,A_{i-1}) \not \in \{-d,-d+1,\dotsc,-1\}$}
\STATE stop the process
\ELSE 
\STATE $G_i \gets $ any gate from $\mathcal{G}^*$ such that $-d \le \diff(G,A_{i-1}) < 0$
\STATE $X_i \gets$ set of variables $G_i$ depends on that are assigned $1$ by $A_{i-1}$
\STATE $y_i \gets$ a uniform random variable from $X_i$
\STATE $A_i \gets$ assignment to $X$ resulting from flipping the value of $y_i$ in $A_{i-1}$
\ENDIF
\ENDFOR
\end{algorithmic}
Clearly, this process decreases the weight of the initial assignment~$A$ by~$1$ at each iteration, for at most~$s$ iterations. In particular, $w(A)-w(A_i)=i$. We now consider three cases.

\noindent \textbf{Case 1.} {\em There exists an assignment $A \in \mathcal{A}^*$ with $\diff(\maj_n,A)=-1$ such that $\walk(A, s, d)$ stops after less than $s$ iterations for some choices of random bits. This means that after $t<s$ iterations, for all the gates $G$ in $\mathcal{G}^*$ we have that either $\diff(G,A_t)<-d$, or $\diff(G,A_t)\geq 0$}

We select randomly a subset $T$ of $t$ variables from $Z=\{x \in X\setminus \{x^*\} \colon A_t(x)=0\}$ and flip them. Denote the resulting assignment by~$A'$. Clearly, $w(A)=w(A')$ and so $\diff(\majn,A')=-1$. 
Therefore there must be a gate $G$ in $\mathcal{G}^*$ such that $-W\leq\diff(G,A')<0$. Thus, before flipping $t$ random variables, all the gates with negative difference has difference less than $-d$, while after the flipping, at least one gate $G$ has difference at least $-W$. Let $Z'=\{x \in X(G) \setminus \{x^*\} \colon A_t(x)=0\}$. This means that the flipping changed the values of at least $r=(d-W)/W$ variables of~$G$, that is, $|T \cap Z'| \ge r$.

Let $p$ be the probability that $|T \cap Z'| \ge r$ where the probability is taken over the random choice of~$T$. By~choosing the parameters $s$ and $d$ we will make $p$ small enough so that with non-zero probability no gate from $\mathcal{G}^*$ satisfies this. Due to the discussion above this leads to a contradiction since flipping $x^\ast$ changes the value of the function, but not the value of the circuit.
The probability that no gate from $\mathcal{G}^*$ satisfies $|T \cap Z'| \ge r$ is at least $1-|\mathcal{G}^*|p$. The probability $p$ can be upper bounded using Lemma~\ref{lem:hypergeo-simple}:
\[p \le \left( \frac{t|Z'|}{|Z|} \right)^r \le \left( \frac{sk}{n/2} \right)^r \,\]
where the second inequality follows since $t<s$, $|Z'|\leq k$ and $|Z| \geq \frac n2$.

We want the probability $1-|\mathcal{G}^*|p$ to be positive. Since $|\mathcal{G}^*| \le k^2/n = n^{1/3+2\eps}$ 
we get the following inequality on $s$, $d$, and~$k$:
$(k^2/n)\cdot (2sk/n)^r < 1 \, .$
We can satisfy this if $sk < n/4$  and $r \geq \log \frac{k^2}{n}$. Since $\log n > \log \frac{k^2}{n}$ for the latter it is enough to have $d = W\log n$.
Overall, this case poses the following constraint for the considered parameters:
\begin{equation}\label{eq:i1}
sk \le n/4 \, .
\end{equation}

\noindent \textbf{Case 2.} {\em For each assignment $A \in \mathcal{A}^*$ (i.e., $\diff(\maj_n,A)=-1$) the process $\walk(A,s,d)$ goes through all $s$ iterations for all choices of random bits.} We consider two subcases here.

\noindent \textbf{Case 2.1.} {\em For each assignment $A \in \mathcal{A}^*$ (i.e., $\diff(\maj_n,A)=-1$) there exists a choice of variables $y_1, \dotsc, y_s$ at line~8 of the process $\walk(A,s,d)$, such that for  each gate $G \in \{G_1, \dotsc, G_s\}$ (recall that the gates $G_1, \dotsc, G_s$ are selected at line~6 of the process) we have $\diff(G,A)\leq f$, where $f$ is again a positive parameter to be chosen later.}

We estimate the expected number $E$ of gates $G$ from $\mathcal{G}^*$ that have $-d\leq\diff(G,A)\leq f$ where the expectation is taken over the random choices of~$A$. Note that a particular gate $G \in \mathcal{G}^*$ may appear in the sequence $G_1, \dotsc, G_s$ at most $d$ times: the first time it appears, it must have $\diff(G,A_1)\leq -1$ for the current assignment $A_1$, the next time it has $\diff(G,A_2)\leq -2$ for the new current assignment $A_2$, and so on. If $Ed < s$ we get a contradiction: take an assignment $A \in \mathcal{A}^*$ with $\diff(\maj_n,A)=-1$ such that the number of gates $G$ in $\mathcal{G}^*$ with $-d\leq\diff(G,A)\leq f$ is at most $E$, then we cannot have that for all of $G_1, \dots, G_s$ it is true that $-d\leq\diff(G_i,A)\leq f$, there are just not enough gates with this $\diff$.

Now we upper bound $E$. Due to the normalization stage any fixed gate has at least $n^2/(100k^2)=n^{2/3-2\eps}/100$ variables in it. Note that the set of inputs $B$ to the gate $G$ that give $\diff(G,B)=i$ for any $i$ form an antichain. Then due to Lemma~\ref{lem:wide-layers-general} the probability for a gate to attain a certain value is at most $O(k/n) = O(1/n^{1/3-\eps})$.

Hence
\[E \le |\mathcal{G}^*|\cdot (f+d)\cdot O\left(\frac{k}{n}\right) = \frac{k^2}{n}\cdot  (f+d) \cdot O\left(\frac{k}{n}\right)= O\left(\frac{k^3 (f+d) }{n^2}\right)= O\left(\frac{k^3f}{n^2}\right)\, ,\]
where for the last equality we add the constraint 
\begin{equation}\label{eq:tr}
d = O(f) \, .
\end{equation}
Overall, this case poses the following constraint for the parameters:
\begin{equation}\label{eq:i2}
O\left(\frac{k^{3}fd}{n^2}\right) = O(fdn^{3\eps})<s \, .
\end{equation}

\noindent \textbf{Case 2.2.} {\em There exists an assignment $A \in \mathcal{A}^*$ (i.e., $\diff(\maj_n,A)=-1$) such that for any choice of variables $y_1, \dotsc, y_s$, for at least one gate $G\in \{G_1, \dotsc, G_s\}$ we have $\diff(G,A)>f$.}

Fix a gate $G \in \mathcal{G}^*$ with $\diff(G,A)>f$.
We are going to upper bound the probability (over the random choices of variables $y_1, \dotsc, y_s$) that $G$ appears among $G_1, \dotsc, G_s$ during the process. If this probability is less than $1/k$, then by the union bound with a positive probability no gate such gate appears among $G_1, \dotsc, G_s$ which leads to a contradiction with the case statement.

For $G$ to appear among $G_1, \dotsc, G_s$, the process has to select a variable appearing in~$G$ at line~8 many times. Indeed, if $G$ appears in the process, then its $\operatorname{diff}$ with the current assignment is negative. At the same time, in the beginning of the process $\diff(G,A)>f$. Each time when the process reduces a variable at line~8 (that is, changes its value from~1 to~0), the value of the linear function computed at~$G$ decreases by at most~$W$ (just because $W$ is the maximum weight of a variable in all the gates in $\mathcal{G}^*$). Thus, it is enough to upper bound the probability that for a fixed gate $G \in \mathcal{G}^*$ with $\diff(G,A)>f$, the process selects a variable from $X(G)$ at least $f/W$ times.

Let $Y_1, \dotsc, Y_s$ be random $0/1$-variables defined as follows: $Y_i=1$ iff the $i$-th reduced variable appears in~$G$ (i.e., $y_i \in X(G)$). 
Let $Y = \sum_{i=1}^s Y_i$.
Our goal is to upper bound $\pr(Y \ge f/W)$.

Let $H_1, \dotsc, H_l$ be all the gates that share at least one variable with~$G$. Assume that on step~$j$ we reduce a variable from~$H_i$. Then
\[\pr(Y_j=1)=\pr(y_i \in X(G))=\frac{|X(G) \cap X(H_i)|}{|\{x \in X(H_i) \colon A_{j-1}(x)=1\}|} \, .\]
Due to the stage~2.1 of the normalization process, $|\{x \in X(H_i) \colon A_{j-1}(x)=1\}| \ge \frac{n^2}{100k^2}-d$. To~see this, assume the contrary. Recall that $-d \le \operatorname{diff}(H_i,A_{j-1})<0$. This means that by increasing at most~$d$ variables (i.e., changing their values from~0 to~1) from $X(H_i)$ in $A_{j-1}$ results in an assignment of weight at most $\frac{n^2}{100k^2}$ that sets $H_i$ to~1. This, in turn, contradicts to the fact that the circuit is normalized. Thus,
\[\pr(Y_j=1) \le \frac{|X(G) \cap X(H_i)|}{\frac{n^2}{100k^2}-d} \le \frac{|X(G) \cap X(H_i)|}{\frac{n^2}{200k^2}}\, ,\]
where we add a constraint
\begin{equation}\label{eq:eqd}
d \le \frac{n^2}{200k^2} \, .
\end{equation}

We are now going to use the fact that variables from a fixed gate $H_i$ can be reduced at most~$d$ times. We upper bound 
$Y=\sum_{i=1}^{s}Y_i$ by the following random variable: 
\[Z=\sum_{i=1}^{l}\sum_{j=1}^{d}Z_{ij} \, .\]
where each $Z_{ij}$ is a random $0/1$-variable such that
\[\pr(Z_{ij}=1)=\frac{|X(G) \cap X(H_i)|}{\frac{n^2}{200k^2}} \,,\]
and $Z_{ij}$ are independent.
That is, instead of reducing variables in some of $H_i$'s in some random order, we reduce~$d$ variables in each~$H_i$. Thus we reduce maximal possible number of variables in all gates. 
Clearly, for any $r$ we have $\pr(Y\geq r) \leq \pr(Z\geq r)$.

Let us bound the expectation of~$Z$. Since due to the normalization each variable of~$G$ appear in other gates at most $100k^3/n^2 = 100n^{3\eps}$ times, we have
\[
\sum_{i,j} |X(G) \cap X(H_i)| \leq d \cdot (100k^3/n^2) \cdot |X(G)|\leq 100\cdot d \cdot k^4/n^2 = 100\cdot n^{2/3+4\eps}\cdot W\cdot \log n.
\]
Overall we get 
\[
EZ \leq\frac{100dk^4/n^2}{n^2/200k^2} = 4\cdot 10^4 \cdot d \frac{k^6}{n^4} =  4 \cdot 10^4 \cdot n^{6\eps} \cdot W\cdot \log n.
\]
Application of Chernoff--Hoeffding bound (Lemma~\ref{lem:chernoff}) immediately implies that the probability that $Z$ is twice greater than the expectation is exponentially small in $d \cdot \frac{k^6}{n^4}$. Since $d \cdot \frac{k^6}{n^4}=W \cdot \log n \cdot n^{9\eps}$ grows asymptotically faster than $\log n$ for sure, we conclude that
\[\pr(Z \ge 2\cdot EZ) < \frac{1}{n} \le \frac{1}{k}\]
Hence, if $f/W \ge 2\cdot EZ$, then
\[\pr(Y \ge f/W) \le \pr(Z \ge 2 \cdot EZ) < \frac{1}{k}\]
as desired. Overall, this gives us the following constraint:
\begin{equation}\label{eq:c22}
f \geq 4 \cdot 10^4 \cdot d \cdot W \cdot \frac{k^6}{n^4} = 4 \cdot 10^4 \cdot n^{9\eps}\cdot W^2 \cdot \log n \, .
\end{equation}

\subsubsection{Tuning the parameters}
It remains to set the parameters so that the inequalities \eqref{eq:i1}--\eqref{eq:c22} are satisfied and $k$ is as large as possible.
The inequality~\eqref{eq:i2} sets a lower bound on $s$ in terms of $f$, while~\eqref{eq:c22} sets a lower bound on~$f$. Putting them together gives a lower bound on~$s$:
\[s \ge 4 \cdot 10^4\cdot \frac{k^9}{n^6} \cdot W^3 \cdot \log ^2 n \, .\]
Combining it with the upper bound on~$s$ from~\eqref{eq:i1}, we can set the following equality on $k$ and~$n$:
\[\frac{n}{4k} = 4 \cdot 10^4\cdot \frac{k^9}{n^6} \cdot W^3 \cdot \log^2 n.
\]
Thus 
\[
k = \Omega\left(\frac{n^{7/10}}{(\log n)^{1/5} W^{3/10}}\right)
\]
and it is easy to see that we with this $k$ we can pick other parameters to satisfy all the constraints
(we set $f$ so that~\eqref{eq:c22} turns into an equality, the inequalities~\eqref{eq:tr} and~\eqref{eq:eqd} are satisfied since $W \leq \frac{k^3}{n^2}$).

 This gives a proof of Theorem~\ref{thm:worst-case-lower-weighted}.
For $W=1$ we get 
\(
k = n^{7/10} \cdot (\log n)^{-1/5},
\)
which gives a proof for Corollary~\ref{cor:worst-case-lower-unweighted}.
For unbounded $W$ recall that we can assume $W \leq \frac{k^3}{n^2}$ and thus 
\(
k = n^{13/19}\cdot (\log n)^{-2/19}
\)
and Theorem~\ref{thm:worst-case-lower-general} follows.

\section{Conclusion and Open Problems} \label{sec:conclusion}
The most interesting question left open is whether one can prove non-trivial upper bounds for $k$ in the worst case. Currently, we do not know how to construct {\model} circuits computing $\maj_n$ on all inputs even for $k=n-2$ (though we have many examples of such circuits for $n=7,9,11$), not to say about $k=n^{\eps}$ for $\eps<1$. 

Another natural open question is to get rid of the logarithmic gap between upper and lower bound for depth-2 randomized circuits.

A natural direction is to extend our studies to the case of non-monotone {\model} circuits.

Many of our results naturally translate to larger depth circuits. Indeed, note that in the proofs of lower bounds we do not use the fact that the function on the top of the circuit is majority. In these proofs it can be any monotone function. Thus we can split a depth-$d$ circuit consisting of $\maj_k$ into two parts: bottom layer and the rest of the circuit. Then our lower bounds translate to this setting straightforwardly. It is interesting to proceed with the studies of larger depth majority circuits.

\section*{Acknowledgments}
We would like to thank the participants of Low-Depth Complexity Workshop (St.~Petersburg, Russia, May 21--25, 2016) for many helpful discussions.

\bibliography{stacs}

\section{Appendix: Omitted Proofs}
\subsection{Technical Lemmas}\label{sec:tech}

\begin{proof}[Proof of Lemma~\ref{lem:wide-layers}]
The probability under consideration is equal to
\[
\Pr[|T\cap S'|=l] = \frac{ \binom{k}{l}\binom{m-k}{t-l}}{\binom{m}{t}}.
\]
It is convenient to introduce notation $c = \frac tm$. Note that then $\eps<c< 1-\eps$. The probability above then can be rewritten as
\[
\Pr[|T\cap S'|=l] = \frac{ \binom{k}{l}\binom{m-k}{cm-l}}{\binom{m}{cm}}.
\]

It is not hard to see that the maximum is achieved for $l$ equal to $ck$ (the probability is increasing for $l < ck$ as a function of $l$ and is decreasing for $l>ck$).

So we need to upper bound 
\begin{equation} \label{eq:prob}
\frac{\binom{k}{ck}\binom{m-k}{c(m-k)}}{\binom{m}{cm}} = 
\frac{\frac{k!}{ck!(1-c)k!}\frac{(m-k)!}{c(m-k)!(1-c)(m-k)!}}{\frac{m!}{cm!(1-c)m!}}.
\end{equation}

To bound the probability we will use Stirling's approximation, the following simple form will be enough 
\[
n! \sim \left(\frac{n}{e}\right)^n \sqrt{n}.
\]
Let us first consider binomial coefficients separately:
\begin{align*}
\frac{m!}{cm!(1-c)m!} &\sim \frac{\left(\frac{m}{e}\right)^m \sqrt{m}}{\left(\frac{cm}{e}\right)^{cm} \sqrt{cm}\left(\frac{(1-c)m}{e}\right)^{(1-c)m} \sqrt{(1-c)m}}\\
& = \frac{1}{(c^{c}(1-c)^{1-c})^m} \cdot\frac{1}{\sqrt{c(1-c)}\sqrt{m}} \\
& = d^m \cdot \frac{1}{\sqrt{c(1-c)}\sqrt{m}},
\end{align*}
where by $d$ we denote $\frac{1}{c^{c}(1-c)^{1-c}}$.

Now for~\eqref{eq:prob} we have
\[
\frac{d^k \cdot \frac{1}{\sqrt{c(1-c)}\sqrt{k}}\cdot d^{m-k} \cdot \frac{1}{\sqrt{c(1-c)}\sqrt{m-k}}}{d^m \cdot \frac{1}{\sqrt{c(1-c)}\sqrt{m}}} = \frac{\sqrt{m}}{\sqrt{c(1-c)}\sqrt{k}\sqrt{m-k}} \sim \frac{1}{\sqrt{k}},
\]
where the last equivalence follows since  $\sqrt{m-k} = \Theta(\sqrt{m})$.

So, we have shown the first part of the lemma and the second part for $l=ck$. To ensure the second part for $|l-ck|<d$ we can compare probabilities for $l$ and $l+1$:
\begin{align*}
\frac{ \binom{k}{l}\binom{m-k}{cm-l}}{\binom{m}{cm}} = 
\frac{ \binom{k}{l+1}\binom{m-k}{cm-l-1}}{\binom{m}{cm}}\cdot
\frac{l+1}{k-l}\cdot \frac{m-k-(cm-l)+1}{cm-l}.
\end{align*}
Note that if $|l-ck|<d$ the probabilities differ by a constant factor.
Thus the asymptotic of the probability is the same for all $l$ satisfying $|l-ck|<d$.
This finishes the proof of lemma.

\end{proof} 

\begin{proof}[Proof of Lemma~\ref{lem:wide-layers-general}]

We introduce the same notation as in the previous proof: $c = \frac tm$. The probability is bounded by
\begin{align*} 
&\frac{ \sum_{r \in A}\binom{m-k}{cm-|r|}}{\binom{m}{cm}} = \sum_{r \in A}\left( \frac{1}{\binom{k}{|r|}}\frac{ \binom{k}{|r|}\binom{m-k}{cm-|r|}}{\binom{m}{cm}} \right) \leq \\
& \max_{|r|} \left( \frac{ \binom{k}{|r|}\binom{m-k}{cm-|r|}}{\binom{m}{cm}} \right) \sum_{r \in A}\frac{1}{\binom{k}{|r|}} \leq 
\max_{|r|} \left( \frac{ \binom{k}{|r|}\binom{m-k}{cm-|r|}}{\binom{m}{cm}} \right),
\end{align*}
where the last inequality is LYM inequality (see e.g. \cite{DBLP:series/txtcs/Jukna11}, Theorem 8.6).

Now we can bound the probability by the same argument as in Lemma~\ref{lem:wide-layers}.

\end{proof}

\begin{proof}[Proof of Lemma~\ref{lem:hypergeo-simple}]
The lemma can be shown by a simple direct calculation:
\begin{multline*}
\pr\{|T \cap Z'| \ge l\} \le 
\frac{\binom{k}{l}\binom{m-k}{t-l}}{\binom{m}{t}} \le k^l \cdot \frac{t}{m} \cdot \frac{t-1}{m-1} \cdot \dotsm \cdot \frac{t-l+1}{m-l+1} \le \\
k^l \cdot \left(\frac{t}{m}\right)^l = \left(\frac{kt}{m}\right)^l,
\end{multline*}
where in the second inequality we use a simple bound $\binom{k}{l}\leq k^l$.
\end{proof}

\subsection{Circuits with High Correlation}
\label{app:corr}

\begin{proof}[Proof of Theorem~\ref{thm:correlation-upper}]
{\em Proof overview.} The required circuit is straightforward: we just pick $k$ random subsets $S_1, S_2, \dotsc, S_k$ of $X$ of size $k$, compute the majority for each of them, and then compute the majority of the results:
\[{\circuit}(X)=\maj_k(\maj_{S_1}(X), \maj_{S_2}(X), \dotsc, \maj_{S_k}(X)) \, .\] 
The resulting circuit has a high probability of error on middle layers of the boolean hypercube. We will however select the parameters so that all the inputs from these middle layers constitute only a small $\varepsilon/2$ fraction. We will then show that 
among all the remaining inputs (not belonging to middle layers) there is only a fraction $\varepsilon/2$ (of all the inputs) where $\maj_n$ may be computed incorrectly. Overall, this gives a circuit that errs on at most $\varepsilon$ fraction of the inputs.

{\em Assignments from middle layers.} Consider all the inputs whose weight differs from $n/2$ by at most $\alpha n^{1/2}$ where $\alpha=\alpha(\varepsilon)$ is a parameter to be chosen later. The number of such inputs is 
\[\sum\limits_{i \colon |i-n/2| \le \alpha n^{1/2}}{n \choose i} \le 2\alpha \cdot n^{1/2}\cdot {n \choose n/2}=2\alpha \cdot n^{1/2} \cdot \Theta(1) \cdot \frac{2^n}{n^{1/2}}=\alpha \cdot \Theta(1)\cdot 2^n \, .\]
By choosing a small enough value for $\alpha=\alpha(\varepsilon)$, one ensures that this is at most $\frac{\varepsilon}{2} \cdot 2^n$.

{\em Assignments from outside of middle layers.} Now, fix an input $A \in \{0,1\}^n$ of weight $n/2+\alpha n^{1/2}$. Pick a random subset $S \subset X$ of size $k=\beta n^{1/2}$ (again, $\beta$ is a parameter to be defined later). We are going to lower bound the following probability (over the choices of~$S$):
\[\pr(\maj_S(A)=1)=\pr(w_S(A) \ge |S|/2)\, .\]
The resulting lower bound will also hold for assignments $A$ of weight greater than $n/2+\alpha n^{1/2}$ (the higher the weight of $A$, the larger is the probability that $\maj_S(A)=1$). By symmetry, it will also give a lower bound on $\pr(\maj_S(A)=0)$ for assignments of weight at most $n/2-\alpha n^{1/2}$.

The distribution of the weight of $A$ on $S$ is a hypergeometric distribution with mean 
\[k \cdot \frac{w(A)}{n}=\beta n^{1/2}/2+\beta\alpha=k/2+\beta\alpha \,. \]
It is known (see, e.g.,~\cite[Corollary~2.3]{DBLP:journals/jal/Siegel01}) that the median of the hypergeometric distribution is approximately equal to its mean.
Hence
\begin{equation}\label{eq:a1}
\pr\left(w_S(A) \ge \lfloor k/2+\alpha\beta \rfloor\right) \ge 1/2 \, .
\end{equation}
By choosing a large enough value of $\beta$, one ensures that $\alpha\beta > 2$. Then Lemma~\ref{lem:wide-layers} guarantees that  
\begin{equation}\label{eq:a2}
\pr\left(k/2 \le w_S(A) < \lfloor k/2+\alpha\beta \rfloor\right) \ge \gamma n^{-1/4}
\end{equation}
for a constant $\gamma>0$.
Collecting~\eqref{eq:a1} and \eqref{eq:a2}, gives us
\[\pr(\maj_S(A)=1)=\pr(w_S(A) \ge k/2) \ge 1/2+\gamma n^{-1/4}\, .\]

Now, pick sets $S_1, S_2, \dotsc, S_k$ of size $k$ uniformly and independently. For each $S_i$, let $Y_i$ be a 0/1-random variable defined by $Y_i=\maj_{S_i}(A)$. Then $\pr(Y_i=1) \ge 1/2+\gamma n^{-1/4}$ and 
\[E\left(\sum_{i=1}^kY_i\right)=k\cdot (1/2+\gamma n^{-1/4})=k/2+\beta\gamma n^{1/4} \, .\]
By Chernoff--Hoeffding bound (Lemma~\ref{lem:chernoff}), the resulting circuit (where the first level gates compute majorities over subsets $S_1, S_2, \dotsc, S_k$) computes $\maj_X(A)$ incorrectly is 
\begin{multline*}
\pr\left(\sum_{i=1}^{k}Y_i < k/2\right)=
\pr\left(\sum_{i=1}^{k}Y_i<E\left(\sum_{i=1}^{k}Y_i\right)-\beta\gamma n^{1/4}\right) \le\\ \exp\left(-\frac{2\beta^2\gamma^2 n^{1/2}}{\beta n^{1/2}}\right)=
\exp(-2\beta\gamma^2) \, .
\end{multline*}
By choosing a large enough value for $\beta$ one makes this expression small enough.

Thus, there exists a choice of $S_1, S_2, \dotsc, S_k$ such that the fraction (among all $2^n$ inputs) of all the inputs from outside of middle layers for which the corresponding circuit computes $\maj_X$ incorrectly is at most $\varepsilon/2$. This gives a circuit that computes $\maj_X$ correctly for at least a fraction $(1-\varepsilon)$ of all the inputs.
\end{proof}

\begin{proof}[Proof of Theorem~\ref{lem:corrlow}]
Let $k=\alpha n^{1/2}$ for a parameter $\alpha=\alpha(\varepsilon)$ to be chosen later. We are going to show that one can set this parameter so that a {\model} circuit errs on more than a fraction $\varepsilon$ of inputs. Note that such a circuit can read at most $k^2=\alpha^2n$ of the input bits. 
Let $R$ be the input bits that are read by the circuit ${\circuit}$ and $U=X \setminus R$ be all the remaining input bits (for read and unread). Then $|R| \le \alpha^2n$. Intuitively, when $\alpha$ is small, the circuit does not even read a large fraction of input bits and for this reason errs on a large number of inputs. We formalize this intuition below.

If $|R|< \alpha^2n$ it is convenient to extend $|R|$ to $|R|=\alpha^2n$, so that $|U|=(1-\alpha^2)n$ and the circuit ${\circuit}$ reads only some of the input bits from $R$ and does not read any input bits from~$U$. Let $\beta$ be a parameter to be chosen later. Denote by $C_R$, $F_R$, $C_U$, $F_U$ the set of all assignments to the variables from $R$ and $U$, respectively, whose weight is close to or far from the middle value, respectively:
\begin{align*}
C_R = \{A \colon R \to \{0,1\} \mid |w(A)-|R|/2| \le \beta n^{1/2}\},\quad & F_R =\{A \colon R \to \{0,1\} \mid A \not \in C_R\},\\
C_U = \{A \colon U \to \{0,1\} \mid |w(A)-|U|/2| \le \beta n^{1/2}\},\quad & F_U = \{A \colon U \to \{0,1\} \mid A \not \in C_U\}.
\end{align*}
We would like to set the parameters $\alpha$ and $\beta$ so that both $|F_U|$ and $|C_R|$ are large enough. Namely, that each of them has at least a fraction $1-\varepsilon/10$ of all the corresponding assignments.

By Lemma~\ref{lem:chernoff}, for a randomly chosen assignment $A \colon R \to \{0,1\}$,
\begin{equation}\label{eq:fr}
\pr(A \in F_R) \le \exp\left(-\frac{2\beta^2n}{|R|}\right)=\exp\left(-\frac{2\beta^2}{\alpha^2}\right) \,.
\end{equation}
On the other hand,
\begin{equation}\label{eq:cu}
|C_U|=\sum\limits_{i \colon |i-|U|/2|\le \beta n^{1/2}}{|U| \choose i}\le 2\beta \cdot n^{1/2} \cdot {|U| \choose |U|/2}
=2^{|U|}\cdot \Theta(1)\frac{\beta}{(1-\alpha^2)^{1/2}}
\end{equation}
We now tune the parameters. First, set $\beta=\frac{\alpha}{\sqrt{2}}\ln\frac{10}{\varepsilon}$ to ensure that~\eqref{eq:fr} is at most $\varepsilon/10$.
Then one can choose a small enough value for $\alpha$ so that~\eqref{eq:cu} is also at most $2^{|U|}\cdot \varepsilon/10$. This is possible, since the function $\frac{\alpha}{(1-\alpha^2)^{1/2}}$ decreases to $0$ with $\alpha \to 0$.

Now, break assignments from $F_U$ into pairs: $A$ and $\neg A$ (clearly, if the weight of $A$ is far from the middle, then so is the weight of $\neg A$, since $w(A)=|U|-w(\neg A)$). Consider an assignment $A \in F_U$, its mate $\neg A \in F_U$, and an assignment $B \in C_R$. Consider the following two assignments to~$X$: $A \sqcup B$ and $\neg A \sqcup B$. Clearly, 
\[\maj_X(A \sqcup B) \neq \maj_X(\neg A \sqcup B) \, .\]
On the other hand, the circuit ${\circuit}$ outputs the same for both of them as it only reads the bits from~$R$. This means that it errs on at least one of these two assignments. This, in turn, implies that the circuit errs on at least a fraction $(1-\varepsilon/10)^2$ of all $2^n$ assignments. For $\varepsilon \le 1/3$, this is grater than $\varepsilon$, a~contradiction.

\end{proof}

\subsection{Randomized Circuits}
\label{app:randc-proof}

\begin{proof}[Proof of Lemma~\ref{lem:randc}]
Consider a randomized circuit {\randcircuit}.
For any minterm/maxterm~$A$
of $\maj_n$, the circuit {\randcircuit} computes $\maj_n(A)$ correctly with probability at least $1-\eps$. This means that one can pick a deterministic circuit {\circuit} from {\randcircuit} that computes $\maj_n$ correctly on at least a fraction $1-\eps$ of all minterms and maxterms of $\maj_n$.

For the other direction, consider a circuit {\circuit} computing $\maj_n$ correctly on at least $1-\eps$ fraction of minterms and maxterms. Let $t={n \choose n/2}$ be the number of minterms. Then we also have $t$ maxterms (for this, we assume additionally that $n$ is odd). The circuit {\circuit} errs on at most $2t\eps$ of minterms/maxterms. 
Consider a random permutation of inputs of~{\circuit}. Denote the resulting distribution of the circuits by~{\randcircuit}. Consider a minterm~$A$ (the case of maxterms is handled similarly). It is not difficult to see that for a randomly and uniformly chosen permutation of its coordinates one gets a uniformly distributed random minterm. Note the the fraction of errors of {\circuit} among minterms is at most $2t\eps/t=2\eps$. Hence {\randcircuit} is incorrect on $A$ with probability at most~$2\eps$.

Now, consider an arbitrary assignment $A \colon X \to \{0,1\}$ such that $\maj_n(A)=1$ (again, the case $\maj_n(A)=0$ is handled in a similar fashion). Then there is a minterm $A' \colon X \to \{0,1\}$ such that $\maj_n(A')=1$ and $A' \le A$ (componentwise). The randomized circuit {\randcircuit} is incorrect on $A'$ with probability at most $2\eps$. Since {\randcircuit} is monotone it is also incorrect on $A$ with at most the same probability.
\end{proof}

\begin{proof}[Proof of Theorem~\ref{thm:randupper}]
Let $p,t$ be parameters to be chosen later. Partition the set of $n$ input bits into $\frac np$ blocks of size $p$: $X=X_1 \sqcup X_2 \sqcup \dotsc \sqcup X_{\frac np}$. For each block $X_i$, compute $[\sum_{x \in X_i}x \ge m]$ for all $m \in [p]$. The outputs of all these $p$ gates is just a permutation of $X_i$, that is, $X_i$ in sorted order. Computing the majority of all these gates (for all blocks) gives us a depth two formula computing $\majn(X)$ with the fanin of the output gate equal to~$n$. To reduce this fanin, instead of going through all values of $m\in [p]$ we go only through $t$ middle values. Thus, the resulting formula looks as follows: on the bottom level, for each block $X_i$, we compute $[\sum_{x \in X_i}x \ge m]$ for all $m \in [\frac p2 - \frac t2..\frac p2 + \frac t2]$; on the top level we compute the majority of all the gates from the bottom level. The fanin of the bottom level of the resulting formula is~$p$ while its top level fanin is $\frac{nt}{p}$. Hence, for this formula
\begin{equation}\label{eq:fanin}
k = \max\left\lbrace p, \frac{nt}{p} \right\rbrace \, .
\end{equation}
A~simple observation is that, if for an assignment $A \colon X \to \{0,1\}$,
\begin{equation}\label{eq:t}
\frac p2 -\frac t2 \le \sum_{x \in X_i}A(x) \le \frac p2 +\frac t2
\end{equation}
for all~$i$, then our formula outputs $\majn(A)$ on the input assignment~$A$.

We turn to estimating the number of assignments $A$ satisfying~\eqref{eq:t}. The number of assignments to $X_i$ violating \eqref{eq:t} is at most
\[2\cdot \sum_{m > \frac p2 +\frac t2} {p \choose m} \, .\]
Hence the total number of assignments $A$ for which the formula computes $\majn$ incorrectly is at most
\[O\left(2^{n-p} \cdot \frac np \cdot \sum_{m > \frac p2 +\frac t2} {p \choose m}\right)\]

We are going to set the parameters $p$ and $t$ such that this number is at most $\frac{2^n}{\operatorname{poly}(n)}$. For this, take $p=n^{\frac 23}$ and $t=\alpha\sqrt{p\ln p}=O(n^{\frac 13}\log^{\frac 12} n)$ (where $\alpha$ is a constant) and use the estimate~\eqref{eq:movemid}.
From \eqref{eq:fanin} we conclude that this gives a {\model} circuit with $k=O(n^{\frac 23}\log^{\frac 12} n)$.
\end{proof}

\begin{proof}[Proof of Theorem~\ref{thm:randlow}]
Consider a {\model} circuit {\circuit} computing $\maj_n$ for $k=\alpha n^{2/3}$. We will show that for small enough value of the constant $\alpha$ such a circuit must err on more than $\eps$ fraction of minterms and maxterms.

For a 
function $f \colon \{0,1\}^n \to \{0,1\}$, define its boundary as follows:
\newcommand{\bound}{\operatorname{Bnd}}
\[\bound(f)=\{(A, i) \colon A \in \{0,1\}^n,\, i \in [n], \, f(A) \neq f(A^i)\} \, ,\]
where by $A^i$ we denote an assignment from $\{0,1\}^n$ resulting from $A$ by flipping its $i$-th bit. In particular, by Lemma~\ref{lem:binomials}, $|\bound(\maj_n)|=\Omega(2^n\cdot n^{1/2})$. Below, we show that for small enough value of $\alpha$, $|\bound(\circuit)|$ is much smaller than $|\bound(\maj_n)|$, which implies that {\circuit} errs on a large fraction of minterms and maxterms of $\maj_n$.

Consider $(A,i) \in \bound(\circuit)$. This means that {\circuit} contains a gate $G$ at a bottom level such that $G(A) \neq G(A^i)$. Recall that $G$ is a monotone
function on $l \le k$ variables. It is known (see, e.g.,~\cite[Theorem~2.33]{DBLP:books/daglib/0033652})  that the influence of such a function is $O(l^{1/2})$: \[\operatorname{Inf}(G)=2^{-l}\cdot \sum_{A \in \{0,1\}^l}|\{i \in [l] \colon G(A) \neq G(A^i)\}|=O(l^{1/2})=O(k^{1/2}) \, .\]
Hence, 
\[|\{(A,i) \colon A \in \{0,1\}^l,\, i \in [l], \, G(A) \neq G(A^i)\}|=O(k^{1/2}2^l) \, .\]
Note that by Lemma~\ref{lem:binomials} any $A \in \{0,1\}^l$ such that $G(A) \neq G(A^i)$ can be extended to a minterm/maxterm of $\maj_n$ in $O(2^{n-l}\cdot(n-l)^{-1/2})$ ways. Thus, $G$ contributes at most
\[O(k^{1/2}\cdot 2^n \cdot n^{-1/2})\]
pairs $(A,i)$ to $\bound(\circuit)$ (note that $(n-l)^{1/2}=\Theta(n^{1/2})$ since $l \le k=\Theta(n^{2/3})$). Since {\circuit} contains at most $k$ such gates, we conclude that
\[\bound(C)=O(k^{3/2}\cdot 2^n \cdot n^{-1/2}) \, .\]
For small enough constant $\alpha$,
\[\bound(C) \le \frac{1}{100} \cdot \frac{n}{2} \cdot {n \choose n/2} \, .\]

In particular, there are at most $\frac{1}{10}\binom{n}{n/2}$ maxterms that contribute at least $n/10$ elements to $\bound(\circuit)$. Thus there are at least $\frac{9}{10}\binom{n}{n/2}$ maxterms that contribute to $\bound(\circuit)$ less than $n/10$ elements. Since by our assumption {\circuit} computes $\maj_n$ correctly on at least $8/10$ fraction of maxterms we have that there is a set $M$ of at least $\frac{1}{2} \binom{n}{n/2}$ maxterms on which the computation of {\circuit} is correct, but the contribution to $\bound(\circuit)$ is small. That is, $M$ consists of assignments $A \colon X \to \{0,1\}$ such that there are at least $4n/10$ of $i$'s for them with $A_i=0$, $(A,i) \notin \bound(\circuit)$, and $\circuit(A)=0$. From this we will deduce that {\circuit} computes $\maj_X$ incorrectly on a large fraction of minterms.

Indeed, consider the following bipartite graph. The vertices of one part are elements of~$M$. For each $A\in M$ and for each $i \in [n]$ with the properties above there is an outgoing edge corresponding to this pair $(A,i)$. The other endpoint of this edge is labeled by $A^i$. Note that $A^i$ is a minterm of $\maj_n$ and by the analysis above ${\circuit}(A^i)=0$. The vertices on the second part of the graph are thus labeled by minterms connected to maxterms in~$M$. It is left to estimate the number of elements in the second part. For this note that there are at least $\frac{1}{2} \binom{n}{n/2}$ vertices in $M$ each of degree at least $4n/10$. On the other hand the degree of each vertex in the second part is at most $n/2$. From this it follows that there are at least 
$$
\frac{1}{2} \cdot \binom{n}{n/2} \cdot \frac{4n}{10} \cdot \frac{2}{n} = \frac{4}{10}\cdot \binom{n}{n/2}
$$ 
vertices in the second part. Thus, the circuit {\circuit} gives the wrong output on at least $4/10$ of minterms, a~contradiction.

\end{proof}

\subsection{Deterministic Circuits}\label{app:det}
\begin{proof}[Proof of Lemma~\ref{lem:n-2-lower}]
Suppose $n=2l+1$ and suppose there is a depth-2 circuit $F$ computing $\maj_n$, consisting of standard majorities of exactly $2l-1$ variables each and for each gate on the bottom layer having distinct variables as its inputs.

Consider the following undirected graph $G$. Its vertices are the inputs $x_1, \ldots, x_n$. Two vertices $x_i$ and $x_j$ are connected if there is a gate on the bottom layer that gets on input all variables except $x_i$ and $x_j$. Thus, graph $G$ has $n$ vertices and $n-2$ edges.

Consider a minterm $A$ of the function $\maj_n$. Its weight is $w(A)=l+1$. For the circuit $F$ to output $1$ on $A$ there should be at least $l$ gates on the bottom layer outputing $1$ on $A$. For each of these gates to output $1$ it has to receive at least $l$ ones on inputs. This is equivalent to saying that one of the two variables that are not given on the input of the gate should be $0$.

Thus in terms of the graph $G$, for the circuit to compute the function correctly it is needed that for any coloring of $l$ vertices of $G$ in color $0$ there are at least $l$ edges that have an endpoint colored in $0$. It is not hard to see that this is impossible. Below we provide a formal proof.

We will construct a coloring of $l$ vertices into color $0$ such that there are at most $l-1$ edges having an endpoint colored in $0$.
Since $G$ has $n$ vertices and $n-2$ edges we have that there are at least two connected components in $G$. For each connected component $H$ consider the following parameter: $p(H) = e(H)-v(H)$, where $v(H)$ and $e(H)$ are the number of vertices and the number of edges in $H$ respectively. The sum of $p(H)$ over all components of $G$ is equal to $-2$. The minimal possible value of $p(H)$ is $-1$ (when $H$ is a tree). Thus, there are at least two components $H$ with negative $p(H)$, that is with $p(H)=-1$. At least one of these components has at most $l$ vertices.
Order the components in the increasing order of the parameter $p(H)$. Among components with $p(H)=-1$ order the component in the increasing order of the number of vertices. Thus the first component is always a tree of size at most $l$.

Now we are ready to color $l$ vertices of the graph in the color $0$. We color all vertices in the first several components and if needed we will color a part of one more component.

If after coloring $l$ vertices we colored completely several components and have not started the next one, then clearly the sum of $p(H)$ over colored components is negative and thus the number of edges with an endpoint colored in $0$ is less than $l$.

Suppose we have colored several components and we need to color a part of the next component $H$. We will explain now how to do it. If $p(H)=-1$, then $H$ is a tree. Color a part of  $H$ of needed size in such a way that the number of vertices in $H$ colored in $0$ is the same as the number of edges with an endpoint colored in $0$ (for example, we can repeat the following procedure: color a leaf and remove it from the tree). Note that in the previous components the sum of the parameters $p$ is negative and we are done. If $p(H)=m\geq 0$ then the sum of parameters $p$ of all colored components is at most $-m-2$. Consider a spanning tree of $H$. It is obtained from $H$ by removing $m+1$ edges. Color a part of the spanning tree of $H$ in such a way that the number of colored vertices in the spanning tree is the same as the number of edges with an endpoint colored in $0$. If we return edges removed from $H$ it will add at most $m+1$ edges with an endpoint colored in $0$. However, in all components in total the number of vertices colored in $0$ is still greater than the number of edges with an endpoint colored in $0$. Thus we have constructed a needed coloring and thus found an input on which the circuit gives the wrong output.
\end{proof}

\begin{proof}[Proof of Theorem~\ref{thm:depth-3-upper}]
We adopt the strategy of the proof of Theorem~\ref{thm:randupper}. That is, we break inputs into $O(n^{1/3})$ blocks, compute majorities on each block on middle $O(n^{1/3})$ layers and then compute the majority of the results. We use the third layer of majority gates to induce additional structure on the inputs. 

We proceed to the formal proof. Partition the set of inputs into $b=n^{1/3}/2^{1/3}$ blocks of size $p=2^{1/3}n^{2/3}$ each: $X=X_1 \sqcup X_2 \sqcup \dotsc \sqcup X_{b}$. For each block $X_i$, compute $[\sum_{x \in X_i}x \ge k]$ for all $k \in [p]$. This constitutes the first layer of the circuit. The outputs of each of these $p$ gates is just a permutation of $X_i$, that is, $X_i$ in decreasing order.

As an output of the first layer we have again $n$ bit vector $Y$ with the same number of $1$'s and $0$'s as in the input, but in each block the bits are ordered in decreasing order. On~the second layer we split $Y$ again into $b$ blocks of size $p$: $Y=Y_1 \sqcup Y_2 \sqcup \dotsc \sqcup Y_{b}$. But now block $Y_i$ consists of the bits of $Y$ with numbers $i, i+b, i+2b, \ldots, i+(p-1)b$. For each block $Y_i$, we compute $[\sum_{y \in Y_i}y \ge k]$ for all $k \in [\frac p2 - (\frac{n}{2})^{1/3}..\frac p2 + (\frac{n}{2})^{1/3}]$. Thus on the second layer we have $2^{2/3}n^{1/3}$ gates for each of $b=n^{1/3}/2^{1/3}$ blocks, that is $2^{1/3}n^{2/3}$ outputs in total. Finally, on the third level we compute the majority of all of the outputs on the second layer.

Now we need to show that this circuit computes the majority for all possible inputs. Since both the circuit and the majority function are monotone, it is enough to ensure that the computation is correct on min-terms and max-terms of majority. 

Consider an input $A \colon X \to \{0,1\}$ with $w(A)=n/2$. We will show, that for each block~$Y_i$,
\begin{equation}\label{eq:rt}
w_{Y_i}(A) \in \left[\frac p2 - \left(\frac{n}{2}\right)^{1/3}, \, \frac p2 + \left(\frac{n}{2}\right)^{1/3}\right] \, .
\end{equation}
Indeed, since the variables in each $X_i$ are ordered and we include in~$Y_i$ each $b$-th variable of each~$X_j$, 
\[w(A) \in [w_{Y_i}(A)\cdot b - b^2, w_{Y_i}(A)\cdot b + b^2], \]
where in $\pm b^2$ the first $b$ factor corresponds to the error in each block $X_i$ and the other $b$ factor corresponds to the number of blocks $X_1,\ldots, X_b$.
On the other hand, we know that $w(A)=n/2$. Thus
\[\frac n2 \in [w_{Y_i}(A)\cdot b - b^2, w_{Y_i}(A)\cdot b + b^2] \] which implies~\eqref{eq:rt}.
Now,~\eqref{eq:rt} implies that the computation of the constructed circuit on $A$ is correct. Indeed, by~\eqref{eq:rt}, on the block $Y_i$, the assignment $A$ has at least $(\frac p2 -b)$ zeroes and at least $(\frac p2 -b)$ ones. This, in turn means that by computing $[\sum_{y \in Y_i}y \ge k]$ only for middle values of~$k$ (namely, $k \in [p/2-b, p/2+b]$), but not for all $k \in [p]$, preserves a balance between 0's and 1's:
\[\maj\left(\left\{[\sum_{y \in Y_i}A(y) \ge k]\right\}_{k \in [p]}\right) = \maj\left(\left\{[\sum_{y \in Y_i}A(y) \ge k]\right\}_{k \in [p/2-b, p/2+b]}\right)\, .\]
\end{proof}

\end{document}